\documentclass[aps,prb,twocolumn,showpacs]{revtex4}

\usepackage{latexsym}
\usepackage{times}
\usepackage{graphicx}
\usepackage{amsmath}
\usepackage{color}

\newcommand{\cao}{{CoAl$_2$O$_4$}}
\newcommand{\mss}{{MnSc$_2$S$_4$}}

\begin{document}
\date{\today}

\title{Impurity Effects in Highly Frustrated Diamond Lattice Antiferromagnets}

\author{Lucile Savary}
\affiliation{Ecole Normale Sup\'{e}rieure de Lyon, 46, all\'{e}e d'Italie, 69364 Lyon Cedex 07}
%\affiliation{AWOL insitute of theoretical Physics}
%\affiliation{Department of Physics, University of California, Santa Barbara, CA 93106}

\author{Emanuel Gull}
\affiliation{Department of Physics, Columbia University, New York, NY 10027}

\author{Simon Trebst}
\affiliation{Microsoft Research, Station Q, 
University of California, Santa Barbara, CA 93106}

\author{Jason Alicea}
\affiliation{Department of Physics and Astronomy, University of California, Irvine,
  CA 92697}

\author{Doron Bergman}
\affiliation{Department of Physics, California Institute of Technology, Pasadena, CA 91125}

\author{Leon Balents}
\affiliation{Kavli Institute for Theoretical Physics, University of California, Santa Barbara, CA 93106}

\begin{abstract}
  We consider the effects of local impurities in highly frustrated
  diamond lattice antiferromagnets, which exhibit large but
  non-extensive ground state degeneracies.  Such models are appropriate
  to many A-site magnetic spinels.  We argue very generally that
  sufficiently dilute impurities induce an {\sl ordered} magnetic ground
  state, and provide a mechanism of degeneracy breaking.  The states
  which are selected can be determined by a ``swiss cheese model''
  analysis, which we demonstrate numerically for a particular impurity
  model in this case.  Moreover, we present criteria for estimating the
  stability of the resulting ordered phase to a competing frozen (spin
  glass) one.  The results may explain the contrasting finding of frozen
  and ordered ground states in \cao\ and \mss, respectively.
\end{abstract}
\pacs{75.10.Jm, 75.10.Pq}
\maketitle

\section{Introduction}
\label{intro}
A common feature of highly frustrated magnets is the existence of a
large (classical) ground state degeneracy in model
Hamiltonians.\cite{Moessner:06} Although this degeneracy is accidental,
in the sense that the multitude of ground states are generally not
symmetry-related, it nevertheless yields striking physical consequences.
For instance, over a broad temperature range the system resides in a
``cooperative paramagnetic'' or ``classical spin liquid'' regime, where
the spins avoid long-range order but fluctuate predominantly within the
ground state manifold.  The ultimate fate of such highly frustrated
spins at the lowest temperatures poses an interesting and experimentally
important problem.  Typically, at very low temperatures entropic or
quantum fluctuations alone are sufficient to lift the degeneracy and
produce an ordering transition via ``order by
disorder''.\cite{Villain:80,Henley:89} However, additional weak effects
which would otherwise be negligible in unfrustrated systems---such as
small further-neighbor exchange\cite{Bergman06,nature07}, spin-lattice
coupling\cite{Yamashita00,Bergman06}, and dipolar
interactions\cite{Bramwell:01}---can also provide a degeneracy-lifting
mechanism, which indeed often dominates over fluctuation effects.

In this paper we discuss degeneracy breaking by quenched random
impurities. Generally even a non-magnetic defect (i.e. one which does
not break spin-rotational symmetry) such as a random bond, an interstitial
spin, or a vacancy, will locally distinguish the various degenerate states
of the pristine system. This brings up a number of issues. First, can
impurities consequently lead to ordering, \emph{i.e}.\ ``order by quenched
disorder''? Or, by virtue of their randomness, do they lead instead to
a glassy disordered state? Do these impurities influence the spins in their vicinity independently from one another?  
Or are their effects rendered highly coordinated by the correlated nature of
fluctuations in the cooperative paramagnetic regime? 

The answers to these questions probably depend in detail upon the nature
of the magnetic system under consideration, particularly the degree of
frustration.  Generally, with increasing frustration comes increasing
ground state degeneracy. One often useful characterization scheme for
frustration involves counting the distinct magnetic ordering wavevectors
which are possible within the classical ground state manifold. In mildly
frustrated magnets, such as the nearest-neighbor triangular
antiferromagnet, this wavevector is unique. In the nearest-neighbor fcc
antiferromagnet, the ordering wavevectors form continuous
one-dimensional lines.\cite{fcc:05} The much more frustrated
nearest-neighbor kagome and pyrochlore antiferromagnets, by contrast,
have ordering wavevectors that fill all of reciprocal
space.\cite{Chalker92,Reimers91}

In the latter kagome and pyrochlore cases, the degeneracy is {\sl
  local}---\emph{i.e}.\ the ground state entropy is extensive, and
states within the ground state manifold are related by modifications of
only a small number of spins.  An impurity can then fix the spin
configuration in its neighborhood, while constraining the spins outside
of its vicinity very little.\cite{Shender93} Since each random impurity
fixes a spin configuration in its neighborhood, roughly independently of
the others, one may expect as a result a globally random ground state,
\emph{i.e}.\ a spin glass.  In fact, the $T>0$ dynamics of such
defective pyrochlore and kagome systems is rather subtle, and the actual
spin glass freezing {\sl temperature} can sometimes be highly suppressed as a
result.\cite{Shender93}  Nevertheless, spin glass behavior is very
commonly observed in highly frustrated magnets,\cite{Ramirez94} even
when the disorder is nominally very weak.

For the other classes of frustrated systems noted above, in which the
ground state ordering wavevectors occupy a smaller subset of reciprocal
space, the degeneracy is sub-extensive.  An infinite number of spins
must then be varied in order to transform one ground state to another.
Thus, different impurities \emph{cannot} independently determine their
local environments.  In this paper, we develop a formalism for dealing
with their effects, focusing for concreteness on the most degenerate
case (of which we are aware) of a sub-extensive degeneracy: frustrated
diamond lattice antiferromagnets.  In a $J_1-J_2$ model on the diamond
lattice, the ordering wavevectors (for antiferromagnetic $J_2>|J_1|/8$)
form a 2d surface within the 3d momentum space.\cite{nature07} This
example is of particular recent interest due to its relevance to the
A-site magnetic spinel materials, with chemical formula AB$_2$X$_4$, in
which magnetic A sites form a diamond sublattice with non-magnetic B and
X atoms.\cite{Loidl:prl04,Loidl:prb05,Loidl:prb06} Because it represents
an extreme case of sub-extensive degeneracy, we expect that the
conclusions obtained for this case apply fairly generally to other less
degenerate frustrated magnets.

Our conclusion is that, for this class of systems, despite the large
ground state degeneracy, long-range magnetic order is stabilized---and
indeed a specific ground state is selected---at sufficiently low
impurity concentrations.  Each impurity induces a small, finite region
around it in which the spins are deformed from an ideal spiral pattern,
like holes in ``swiss cheese'' (Emmentaler).  The swiss cheese model allows a
calculation of the global ground state wavevector, based on certain
properties of an individual defect.  We calculate this wavevector for
the A-site spinel case, with a specific impurity model.  We show how the
same theoretical framework determines other physical properties such as
the ordered moment observed in neutron scattering, and the transition
temperature.  The swiss cheese model also signals its own demise, in one
of two ways.  First, if the holes in the cheese strongly overlap, the
assumption of their independence fails.  Second, even when the holes do
not overlap, if the underlying ``stiffness'' of the bulk spiral is too
small, then the impurities may induce strong fluctuations.  In either
case, the long-range order is expected to give way to a disordered spin
glass ground state.  These two possibilities provide criteria, whereby
the stability of the ordered spiral state can be quantitatively
estimated.  In the case of the A-site spinels, we suggest that this
method consistently explains the contrasting glassy and ordered ground
states found in CoAl$_2$O$_4$\cite{2011arXiv1103.0049M,Loidl:prb05} and MnSc$_2$S$_4$\cite{Loidl:prb06,Loidl:prl04,Giri:prb05}, respectively.

The remainder of the paper is organized as follows.  We consider a
single impurity in Sec.\ \ref{sec:single-impurity}. Using a non-linear
sigma model, it is shown quite generally that, on long length scales, a
single defect can generate only small deformations away from a uniform
spiral ground state of the clean system. We then demonstrate via Monte
Carlo simulations that the classical degeneracy is indeed lifted by the
impurity, which favors specific wavevectors along the spiral surface,
thereby providing a mechanism of ``order by quenched disorder''. A
single impurity is further characterized by a length scale $\xi$ (the
size of the hole) outside of which the spins are well-described by a
uniform spiral. In Sec.\ \ref{sec:dilute-impurities}, we extend this
analysis to the case of multiple impurities. There, we discuss the
interplay between impurity and entropic effects, and make quantitative,
verifiable predictions for how $T_c$ varies with impurity
concentration. We conclude in Sec.\ \ref{sec:discussion} with a
discussion of our results in the context of experiments and impurity
effects in other models.

\section{Single impurity}
\label{sec:single-impurity}

In this section, we discuss the physics of a single impurity. First, we
will consider the possibility that the impurity induces a slow variation
of the spins extending over infinite distances. By analyzing the energy
as a function of order parameter variations, we show that this is not
the case. Instead, the deformation of the spins by each impurity is {\sl
  local}, and decays to a uniform spiral as the distance from the defect
increases. We then show that the impurity physics can be characterized
by an impurity energy function, $E_a(\mathbf{q})$, which gives the difference between
the ground state energies of the system with and without a single
impurity of type $a$, under the constraint that far from the impurity
the spins adopt a spiral configuration with wavevector $\mathbf{q}$.  Employing
extensive Monte Carlo simulations we calculate this function numerically
for a specific impurity model.  In order to check the validity of the
swiss-cheese model, we characterize the local region of deformation
around an impurity: we compute locally the $q$-vector from Monte Carlo
realizations of spin configurations. We find that in our simulations,
variations of $q$ are extremely local, and most changes happen within
one unit cell.

\subsection{General considerations}
\label{sec:gener-cons}

Consider an arbitrary local defect, for which the Hamiltonian of the system can only be modified in
a finite vicinity of the impurity (involving only a finite number of spins). 
Also, for simplicity, we will assume the defect is ``non-magnetic'', meaning it preserves the
spin-rotational invariance of the Hamiltonian.

The energy of the system in the presence of the impurity 
then consists of a contribution in the region where the defect has modified
the Hamiltonian, and a contribution from the remainder of the system.
For any spin configuration, the former is finite and the latter contains a leading
term proportional to $V$ and subdominant corrections. By choosing the spin 
configuration equal to that of one of the ground states in the absence of the 
defect, we can make the energy density $E/V=\epsilon_0$ in the large 
$V\rightarrow \infty$ limit equal to that of the pure system, and therefore
the ground states in the presence of the impurity must also achieve this same 
energy density $\epsilon_0$. This implies that spins far from the impurity must 
locally resemble one of the ground states of the pure system. 

\subsubsection{Spiral order parameter }
\label{sec:spir-order-param}

%-----------------------------
\begin{figure}[t]
\includegraphics[width=0.8\columnwidth]{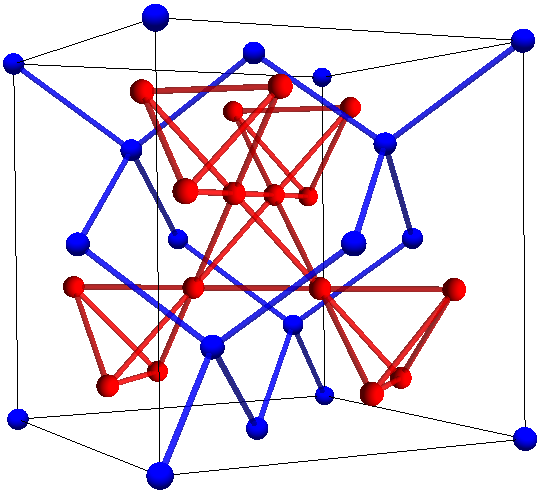}
\caption{
  Cubic cell of an AB$_2$X$_4$ spinel. 
  The sublattice of A sites (blue spheres) is a diamond lattice,
  while the sublattice of B sites (red spheres) is a pyrochlore lattice. 
  }
\label{Fig:Lattice}
\end{figure}
%-----------------------------

To make our discussion more concrete, we now specialize to the case of the 
frustrated diamond lattice antiferromagnet with first and second nearest 
neighbor interactions. The ground states of this system were
determined in Ref.~\onlinecite{nature07}. For $J_2/|J_1|> 1/8$, which is the parameter regime we
focus on hereafter, they consist of coplanar spirals whose propagation
wavevector ${\bf q}$  lies anywhere on a continuous ``spiral
surface'' in reciprocal space. The configuration of the spiral is
described by \begin{equation}
 \label{eq:1}
 \vec{S}({\bf r}) =  {\rm Re}\left[ \vec{d} \,e^{i {\bf q}\cdot{\bf
       r}+i \gamma({\bf q};{\bf r})}\right],
\end{equation}
where the phase $\gamma({\bf q};{\bf
r})=\pm \gamma({\bf q})$ when ${\bf r}$ is on the I or II diamond sublattice,
respectively. (The parametrization is such that the I diamond sublattice contains the site at $(0,0,0)$ and the II sublattice that at $\frac{1}{8}(1,1,1)$).  The vector $\vec{d}$ specifies the plane of the spiral in spin space and its
phase.  It takes the form
\begin{equation}
 \label{eq:2}
 {\vec d} = \hat{e}_1+i\hat{e}_2,
\end{equation}
where $\hat{e}_1,\hat{e}_2$ are orthogonal unit vectors and $|\vec{d}|$
is fixed at $\sqrt{2}$.  The spiral surface itself (i.e. the locus of
allowed ${\bf q}$) deforms smoothly with $J_2/|J_1|$ (except at the
isolated value of $J_2/|J_1|=1/4$ where it changes topology).

To specify a ground state, one must therefore specify both $\vec{d}$ and
the wavevector ${\bf q}$, constrained to the spiral surface.  One can then regard $(\vec{d},{\bf q})$ as the order parameter.  Far from
the impurity, the spin configuration must locally take the ground state
form of Eq.~\eqref{eq:1}, but we must consider the possibility that these
parameters may vary slowly (relative to the largest micro-scale of the
spiral, the wavelength $2\pi/|q|$) in space.  We will now argue that
such variations are insignificant: far from the impurity, the spiral
wavevector and the $\vec{d}$ vector are uniform in the ground state (and
indeed all finite energy states).

To do so, we consider the energy of a slowly-varying order parameter
that is macroscopically non-uniform and show that it is divergent.
Encoding the slow variations na\"ively requires $5$ continuous real
functions: three angles to specify $\vec{d}$, and two more to specify
the position of ${\bf q}$ on the surface.  However, the actual number of
degrees of freedom is smaller due to an additional gauge symmetry: To
see this we note that a change in the wavevector, ${\bf q}\rightarrow
{\bf q}+\delta {\bf q}$ can be compensated by the shift ${\vec
 d}\rightarrow {\vec d}e^{-i \delta{\bf q}\cdot{\bf r} - i \delta\gamma}$ with no change
to the spins (here $\delta \gamma = \gamma({\bf q + \delta q}) - \gamma({\bf q})$). 
Therefore there is a ``gauge'' redundancy in these
variables. We can ``fix'' the gauge in a variety of
ways. A simple choice is to allow {\sl only} for spatial variations in
$d$ and {\sl not} in ${\bf q}$, i.e. we write:
\begin{equation}
 \label{eq:spiral_deviation}
   \vec{S}({\bf r}) =  {\rm Re}\left[ \vec{d}({\bf r}) \,e^{i {\bf
         q}_0\cdot{\bf
         r}+i \gamma({\bf q};{\bf r})}\right] ,
 \end{equation}
 where $\vec{d}({\bf r})$ is assumed to be slowly varying in space, and
 ${\bf q}_0$ is a constant ``reference'' wavevector. We emphasize that
 this still allows the physical wavevector to be different from ${\bf
   q}_0$. For instance, if $\vec{d}({\bf r})=\vec{d}_0 e^{i\delta{\bf
     q}\cdot{\bf r}}$ with constant $\vec{d}_0$, the physical wavevector is ${\bf q}={\bf
   q}_0+\delta{\bf q}$.  In general, we can define the physical wavevector as
 \begin{equation}
 \label{eq:5}
 q^\mu=q^\mu_0 + \frac{1}{2}{\rm Im} \left[ \vec{d}^* \cdot
   \partial_\mu \vec{d}\right] .
\end{equation}
Note that for Eq.~(\ref{eq:spiral_deviation}) to correspond locally to a proper
minimum energy spiral ground state, the first argument ${\bf q}$ of
$\gamma$ must be the physical wavevector given by Eq.~(\ref{eq:5}),
not ${\bf q}_0$.  

\subsubsection{Energy of weakly deformed spirals}
\label{sec:energy-weakly-deform}

It is sufficient to consider just small spatial variations of $\vec{d}$,
since we will find that these are already prohibitively costly at long
distances. Let \begin{equation}
 \label{eq:6}
 \vec{d}({\bf r}) = \vec{d}_0 + \delta \vec{d}({\bf r}).
\end{equation}

To preserve the unit vector constraint of the spins $\vec{S}^2 = 1$ in
Eq.~\eqref{eq:spiral_deviation}, a small $\delta \vec{d}({\bf r})$ must be of the form
\begin{equation}
 \label{eq:7}
 \delta \vec{d}({\bf r}) = i \phi({\bf r}) \vec{d}_0 + \psi({\bf r})
 \hat{e}_3,
\end{equation}
where $\phi$ and $\psi$ are arbitrary small real and complex fields, respectively, and
\begin{equation}
 \label{eq:8}
 \hat{e}_3 = \hat{e}_1 \times \hat{e}_2 = -\frac{1}{2}
{\rm Im} \left[ \vec{d}\times \vec{d}^*\right].
\end{equation}
$\phi$ describes the rotation of the vector $\vec{d}$ within the spiral plane 
(spanned by $\hat{e}_1,\hat{e}_2$),
and includes simple variations in the physical wavevector, while $\psi$ describes
variations outside the spiral plane. To this linearized order, we have simply ${\bf q} = {\bf q}_0+
{\boldsymbol
 \nabla}\phi$.

Now consider the energy density as a function of $\phi,\psi$ and their
gradients.  First, the energy must be unchanged for constant values of
these functions, since these correspond to global $O(3)$ spin rotations.
The first non-trivial terms in a Taylor expansion can arise at quadratic
order in these fields, and from the above reasoning, must include only
spatial gradients so that they vanish for constant
configurations. Finally, this quadratic form must be positive
semi-definite, because the un-deformed configuration obtains the minimal
energy.

An additional constraint is given by frustration: the energy must {\sl
 also} be unchanged for deformations corresponding to changes of the
wavevector within the spiral surface.  Such a deformation is of the form
$\phi({\bf r}) = \delta {\bf q}\cdot {\bf r}$, where $\delta{\bf q}$ is
an arbitrary (small) vector in the plane tangent to the spiral surface
at ${\bf q}_0$.  This constraint is highly restrictive.  Consider the
structure of allowed quadratic terms in $\phi$ with two gradients:
\begin{equation}
 \label{eq:9}
 {\mathcal E}_\phi = \frac{1}{2} c_{\mu\nu} \partial_\mu \phi \partial_\nu
 \phi ,
\end{equation}
where a sum over $\mu,\nu$ is implied, and $c_{\mu\nu}$ is an arbitrary
real symmetric matrix.  This energy density should vanish for a
deformation corresponding to a constant spiral with a wavevector shifted
slightly within the spiral surface, which implies
\begin{equation}
 \label{eq:10}
 c_{\mu\nu}\delta q^\mu \delta q^\nu = 0,
\end{equation}
for $\delta {\bf q}$ in the tangent plane.  Eq.~\eqref{eq:10} reduces $c_{\mu\nu}$ to a single
undetermined coefficient $c$, such that $c_{\mu\nu}= c \hat{n}_\mu \hat{n}_\nu$, where $\hat{\bf
n}$ is the unit normal vector to the spiral surface.  The energy cost to deform $\phi$ in the
directions parallel to the spiral surface is thus higher order in derivatives.  Along the same lines
one may deduce the most general allowed energy density quadratic in the $\phi,\psi$
fields with the minimal number of gradients to ensure stability:
\begin{eqnarray}
 \label{eq:12}
 {\mathcal E} & = & \frac{c}{2} (\nabla_\perp \phi)^2 + c'
 \nabla_\perp \phi \nabla^2_\parallel \phi +
 \frac{c''}{2}(\nabla^2_\parallel \phi)^2 \nonumber \\
 && + d \nabla_\perp \psi^* \nabla_\perp \psi^{\vphantom*} + d'
 {\boldsymbol \nabla}_\parallel \psi^* \cdot {\boldsymbol
   \nabla}_\parallel \psi^{\vphantom*},
\end{eqnarray}
where $\nabla_\perp \equiv \hat{n}\cdot{\boldsymbol \nabla}$,
${\boldsymbol \nabla}_\parallel = {\boldsymbol \nabla}-\hat{\bf n}
\nabla_\perp$, and $c,c',c'',d,d'$ are undetermined coefficients.  For
the energy to be bounded by the ground state value, one needs
$c,c'',d,d'>0$, and $(c')^2 \leq cc''$.  To simplify Eq.~\eqref{eq:12},
we have actually assumed at least a three-fold rotational symmetry about
the axis of the ordering wavevector ${\bf q}_0$.  In the most general
case, the terms involving ${\boldsymbol\nabla}_\parallel$ should be
replaced by less isotropic forms, e.g. $\nabla_\parallel^2 \rightarrow
g_{\mu\nu} \partial_\mu\partial_\nu$, with $\mu,\nu$ spanning the
tangent directions.  However, such changes do not alter the results of
the analysis at the scaling level we consider in this paper.

Now we estimate the energy cost of a deformation.  Consider first $\psi$, whose energy is
determined by the last two terms in Eq.~\eqref{eq:12}.  The scaling is
fully isotropic ($k_\perp \sim k_\parallel$) as usual for an
ordinary Goldstone mode (phonon or magnon) in three dimensions.  This leads to the conventional
estimate of the energy cost for a ``twist'' in the order parameter: if $\psi$ varies by some
finite amount $\delta \psi$ over a region of size $L$, the energy density is increased by an
amount of order $|\delta\psi|^2/L^2$, which integrates to a total energy of order $|\delta
\psi|^2 \times L$ over the volume of size $L^3$.  Since this grows unboundedly with $L$, such
order one distortions of $\psi$ cost infinite energy in the thermodynamic limit, and cannot be
compensated by any local energy gain.

The energy for twists of $\phi$ (which includes wavevector variations)
is less conventional. Here the scaling is anisotropic: if $\phi$ is
distorted by an amount $\delta\phi$ over a distance $L_\parallel$ in a
direction parallel to the spiral surface, it will typically relax over
a larger distance of order $L_\perp \sim L_\parallel^2$ in the
direction perpendicular to the surface.  This is seen simply by
comparing the powers of derivatives in the first three terms of
Eq.~\eqref{eq:12}.  The energy density for such a deformation is then
$(\delta\phi)^2/L_\parallel^4$, which should be integrated over the
volume $L_\perp L_\parallel^2 \sim L_\parallel^4$ to obtain a total
energy {\sl which does not scale with length}.   Thus deformations of
the phase might occur with $O(1)$ disorder contributions, but there
could be subtleties involving thermal fluctuations and anharmonic
elasticity.\cite{PhysRevA.26.915}

In fact, the preference for uniform {\sl wavevectors} at large distances is stronger than the
above estimate might lead one to believe.  The reason is that since $\delta{\bf q}= {\bf q}-{\bf
q}_0 = {\boldsymbol
 \nabla}\phi$, a wavevector shift $\delta q$ (in the spiral surface)
over a region of size $L_\parallel$ leads already to a large (not $O(1)$) deformation of $\phi$:
$\delta\phi \sim L_\parallel \times (\delta q)$.
Following the prior arguments, one sees that a
variation of the wavevector of $\delta q$ over a region of size $L_\parallel$ costs an energy
$\sim (\delta q)^2 L_\parallel^2$.  Thus while more subtle effects could allow for large scale
variations of $\phi$ (see Eq.\eqref{eq:33} and the corresponding discussion), large scale twists of ${\bf q}$ are certainly energetically forbidden in
the ground state.

\subsection{Characterization of single-impurity effects}
\label{sec:char-single-impur}
The preceding discussion implies quite generally that a single impurity can induce order-one deviations from a uniform spiral only locally.  Nevertheless, such corrections are important to quantify
as they break the large spiral degeneracy present in the pure system (at
zero temperature), leading to rich physics. In the following we explain this degeneracy breaking and characterize the
resulting ground states.

\subsubsection{Single-impurity quantities}
\label{sec:single-impur-quant}

To characterize a single impurity, we examine its effect on the spiral
ground states of the pure system. The simplest and most important quantity is
the minimum energy of the system in the presence of the impurity $E(\mathbf{q})$, 
relative to the minimum energy without the impurity, given that infinitely 
far from the impurity the spins are in a spiral configuration with wavevector $\mathbf{q}$.  Formally, for an impurity $a$, $E_a(\mathbf{q})$ is 
\begin{eqnarray}
E_a(\mathbf{q})& =& \text{energy}(\mathbf{q};\text{with
  impurity})\nonumber \\ &&-\text{energy}(\mathbf{q};\text{without impurity}).
\end{eqnarray}
We only need to consider wavevectors $\mathbf{q}$ on the spiral surface, in which case the locality
arguments above imply that $E(\mathbf{q})$ is finite in the infinite volume
limit. This energy quantifies the splitting of the degenerate spiral states by such impurities.

One may also examine the spatial range of the impurity-induced
deformation.  
%We can do this in a variety of ways.  The simplest is to
%consider the deviations of the spins themselves from an ideal spiral
%configuration (enforced at infinity).  One can always choose a ``spin
%deformation length'' $\xi_s$, such that outside some radius $\xi_s$ from
%the impurity, all spins are within some chosen angle $\theta_0$ from the
%ideal.  Alternatively, 
To this end we can {\sl locally} calculate, at each site ${\bf r}_i$, a local spiral 
wavevector ${\bf q}_i$ from the surrounding spin configuration
and then consider the deviation $\cos{(\delta q)} = {\bf q}_i \cdot {\bf q}/(|q_i||q|)$ from
the wavevector  ${\bf q}$ taken at infinity.  
The local measurement of the spiral wavevector ${\bf q}_i$
is performed by considering a set of neighboring spins on the same
sublattice and fitting to
\begin{equation}
 \vec{S}_i\times\vec{S}_j = \sin ({\bf q}\cdot {\bf r}_{ji})
 \frac{i}{2} \vec{d}\times\vec{d}^* = \sin ({\bf q}\cdot {\bf r}_{ij}) \hat{e}_3,
 \label{Eq:LocalWavevector}
\end{equation}
where ${\bf r}_{ji}={\bf r}_j -{\bf r}_i$ and $\hat{e}_3$ defines the 
spin axis perpendicular to the spiral plane as given in Eq.~\eqref{eq:8}.

With the calculated ${\bf q}_i$ we can then define the wavevector
deformation length $\xi_q$ as the radius outside which the angle between
${\bf q}_i$ and ${\bf q}$ at infinity is less than some angle
$\theta_0$.  A cautionary remark is in order.  The finiteness of these
lengths does {\sl not} mean that the deformation around an impurity
decays exponentially away from it.  Rather it means only that the
deformation decays toward a uniform spiral, reaching a ``good''
approximation of it within length $\xi_{q}$.  However, the approach to
the uniform spiral is expected to be in the form of a power-law rather
than exponential, since there is no gap in the spectrum of normal modes
of the spiral state.

Despite this non-exponential decay, the lengths are significant because
the larger they are, the less local the impurity effects become, and the
more sensitive the system is to disorder.  Specifically, we can no longer
regard the impurities as dilute when their concentration is larger than
$\xi_q^{-3}$.   From the above general scaling arguments, we would
expect $\xi_q$ to be typically of the order of a few lattice spacings,
though it might grow larger near special points in the phase diagram.
To check for this possibility, we consider explicitly the size of the
impurity deformation region in a specific impurity model below, and find
that it remains small throughout the parameter range of interest.

\subsubsection{Specific impurity model}
\label{sec:spec-impur-model}

%-----------------------------
\begin{figure}[t]
\includegraphics[width=0.8\columnwidth]{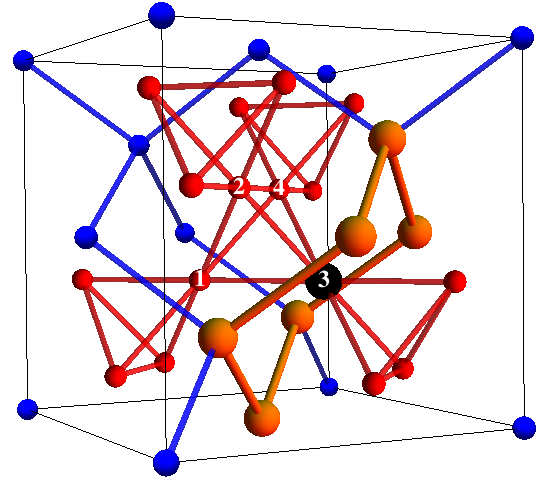}
\caption{
  Impurity model:
  A non-magnetic impurity resides on a B-site indicated by the black sphere.
  The six nearest-neighbor A-sites form a distorted hexagon around the impurity.
}
\label{Fig:Impurity}
\end{figure}
%-----------------------------
In the following we investigate in detail one particular type of impurity
relevant for the spinels, which brings out the general features of the
problem. Specifically, we consider the effect of a magnetic ion on a B
site of the spinel structure AB$_2$X$_4$. Each B-site atom has six nearest-neighbor A sites,
and the distance in this case is smaller than the A-A nearest-neighbor distance. 
Thus, the dominant effect of this impurity is to generate an exchange coupling $J_{\rm imp}$
between the magnetic B-site and its six nearest-neighbor A-sites (see
Figure~\ref{Fig:Impurity}), which is expected to be much stronger than
the A-A exchange, i.e. $J_{\rm imp}\gg J_1,J_2$. We therefore model a
single B-site impurity by adding to the Hamiltonian the term
\begin{equation}
 \delta H = J_{\rm imp}\sum_{\langle a,i\rangle} {\bf S}_a \cdot {\bf S}_i,\label{eq:13}
\end{equation}
where the sum is over the six A-site nearest neighbors $i$ to the B-site
impurity labeled by $a$. Since we expect $J_{\rm imp}\gg J_1,J_2$,
the natural, simplest approximation is to take $J_{\rm imp} \rightarrow \infty$, in which
case the impurity spin ${\bf S}_a$ can be eliminated and Eq.~(\ref{eq:13})
reduces to a boundary condition that the six spins in the vicinity of
the impurity are aligned.

It is noteworthy that the B-site does not have the full point group
symmetry of the lattice. Instead there are four distinct B sites, which
transform into one another under the full set of cubic operations (see
Fig.~\ref{Fig:Impurity}). Therefore we must distinguish the four impurity
positions within the unit cell, which we label $a=1,2,3,4$ in the energy function $E_a(q)$,
as these will favor different ordered states.

\subsubsection{Numerical results}
\label{sec:numerical-results-single-impurity}

%-----------------------------
% floating figure below moved to this place for better placement in output
\begin{figure*}[t]
\includegraphics[width=\linewidth]{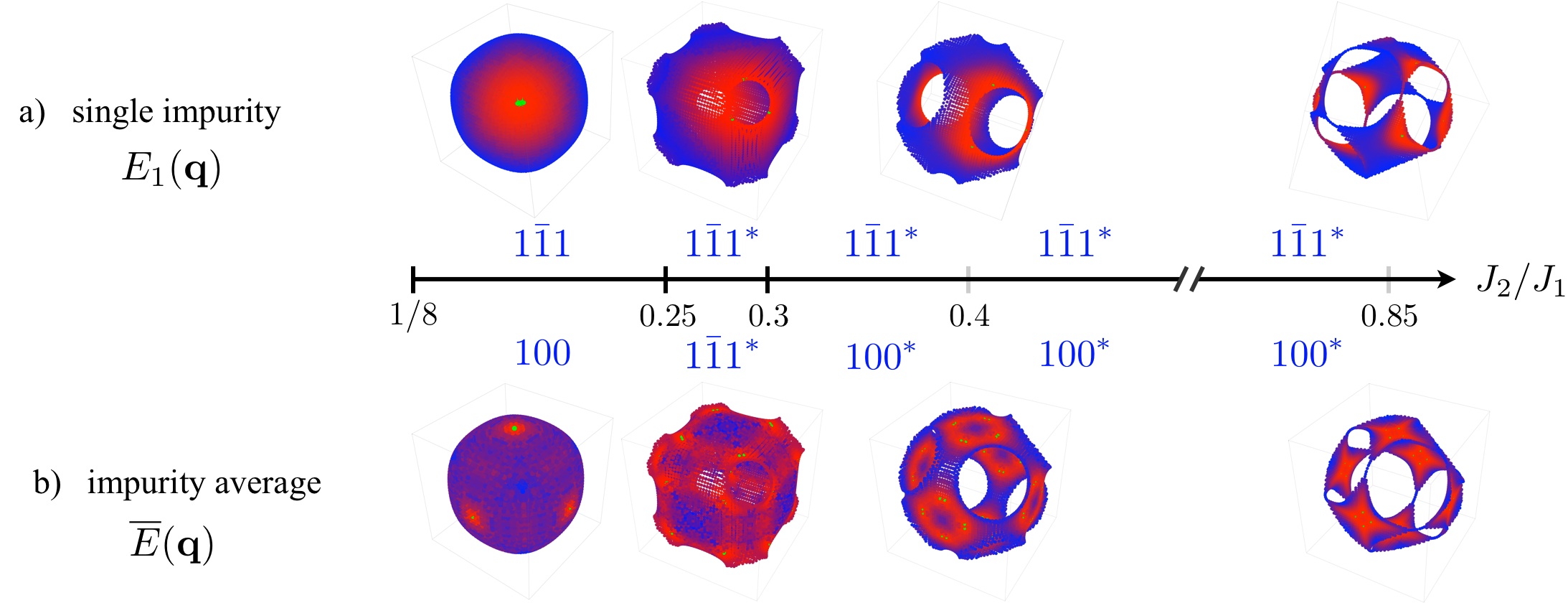}
\caption{
 `Spiral surfaces' comprising the degenerate spiral ground-state vectors
 for varying coupling strengths $J_2/J_1$.
 The surfaces are color-coded according to the energies $E_1({\bf q})$
 and $E({\bf q})$, respectively.  
 The colors indicate high values as blue, low values as red, 
 and green being the absolute minima. 
 The top row shows results for a single impurity $E_1({\bf q})$, 
 while the bottom row shows results averaged over the four possible 
 impurity sites $E({\bf q})$.
}
\label{Fig:ImpuritySelection}
\end{figure*}
%-----------------------------

We have simulated the B-site impurity model numerically by employing
extensive classical Monte Carlo simulations.
We set up our simulations such that the impurity is embedded into
systems of $N=8\times L^3$ spins with system sizes ranging up to $N=8\times 9^3=5832$
spins. 
In order to define the spiral state at large distances from the impurity site 
we employ fixed boundary conditions by embedding the simulation cube of length $L$ 
into a cube of extent $L+1$, where the spins in the boundary layer are aligned to form
a uniform spiral of a given wavevector ${\bf q}$. In the vicinity of the impurity we
consider the $J_{\rm imp} \rightarrow \infty$ limit and force the six nearest-neighbor spins
of the impurity to be aligned and point in the same direction at all times in the simulations.
We explore the zero temperature physics of this impurity model by 
setting the simulation temperature much lower than all energy scales in the problem, 
thereby mimicking a steepest descent energy minimization. 
We checked the convergence of this procedure by simulating systems 
with different initial spin configurations and obtained indistinguishable results
when starting from random spin configurations or unperturbed spiral
states, pointing to the existence of a unique (and well accessible) energy minimum.

Since the four distinct impurity sites within the diamond lattice unit cell
are related by simple rotations, we have calculated the energy $E_a({\bf q})$
only for the impurity at one of these 4 sites. 
%at site ${\bf r}_{a=1} = (3/8,5/8,3/8)$ within the unit cell
%only, a location that is slightly off center. 
%\textcolor{blue}{Doron says: We either need here the lattice vector conventions, basis etc. 
%or simply not specify the cocordinates of this B-site}
For a given value of interactions $J_2/|J_1|$ we have run simulations 
for a set of 1,000 distinct spiral wavevectors ${\bf q}$ on the `spiral surface' appropriate
for the value of couplings. 
A summary of our numerical results for a medium sized system of 
$N=512=8\times4^3$ spins is plotted in the top row of Fig.~\ref{Fig:ImpuritySelection}.
The impurity energies $E_1({\bf q})$ are found to vary on the spiral surfaces and 
clearly reflect the reduced
symmetry of the single B site impurity problem. For instance, in the
coupling range $1/8 \leq J_2/|J_1| \leq 1/4$, where the spiral surface is a 
distorted sphere, the minimum energy wavevectors for $E_1(q)$ are ${\bf q}_1$
points which are along the $1\bar{1}1$ direction, while the energy for wavevectors in the $\bar{1}11$ 
direction (and others in the $\langle111\rangle$ octet) is {\sl not} an energy minimum.

For $J_2/|J_1|> 1/4$, the spiral surface develops `holes' centered around
the $111$ directions and we find that $E_1({\bf q})$ develops three energy minima 
located symmetrically on the spiral surface along the $111^*$ direction 
for all couplings $J_2/|J_1|> 1/4$ as indicated in the top row of 
Fig.~\ref{Fig:ImpuritySelection}.

\begin{figure*}[t]
\includegraphics[width=0.9\columnwidth]{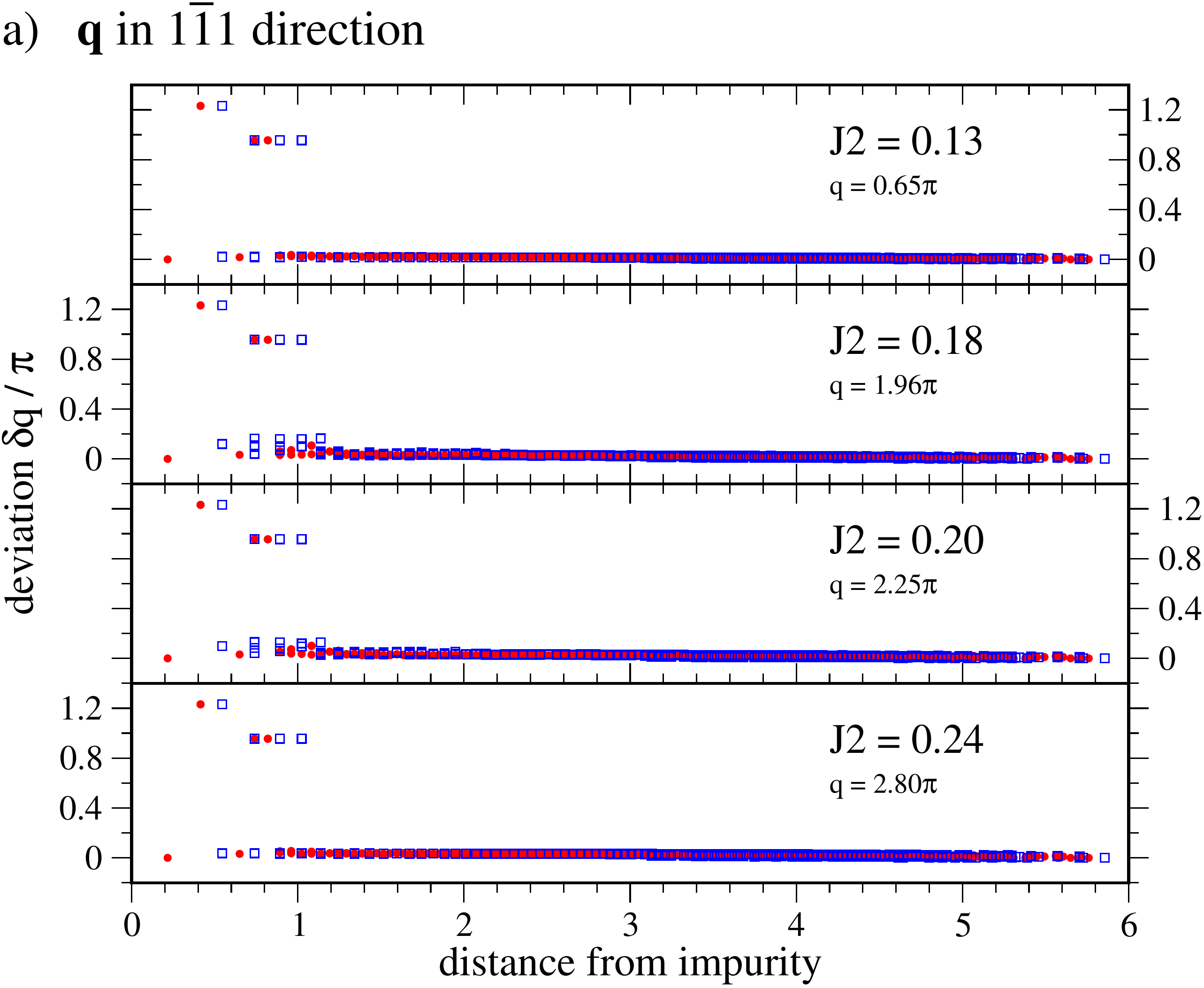}
\hspace{0.18 \columnwidth}
\includegraphics[width=0.9\columnwidth]{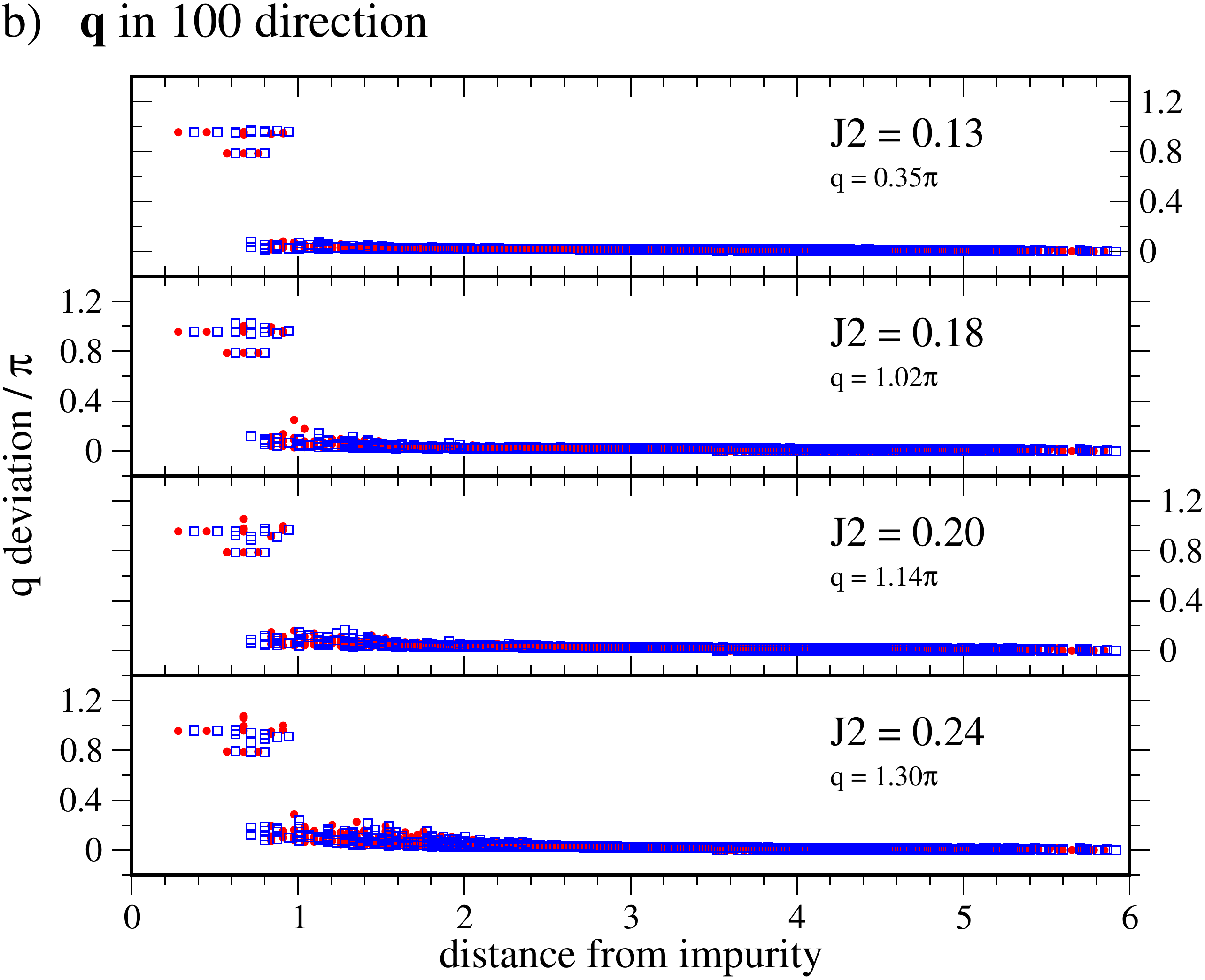}
\caption{
 Deviation of the local spiral wavevector ${\bf q}_i$ from the one of the spiral state 
 at the boundary $\bf q$.
 The left panel is for boundary spiral states with $\bf q$ pointing along 
 the $1\bar{1}1$ direction in the coupling range $J_2/|J_1|<1/4$. 
 The right panel  is for boundary spiral states with $\bf q$ pointing along 
 the $100 $ direction in the coupling range $J_2/|J_1|<1/4$.
 The impurity is embedded into a system of $N=2,744=8\times7^3$ spins.
 The symbols correspond to two different sets of neighboring spins, 
 $P$ (circles) and $Q$ (boxes), used to calculate the local spiral 
 wavevectors, details are given in the text.
}
\label{Fig:SpiralWavevector2_qqq_q00}
\end{figure*}

Our numerical simulations also allow us to probe the spiral deformations in the vicinity of the 
impurity. In particular, we measure the local spiral wavevector ${\bf q}_i$ as 
described in detail in section \ref{sec:single-impur-quant} using 
Eq.~\eqref{Eq:LocalWavevector}. 
Since we are mostly interested in estimating the deviation of this local
wavevector ${\bf q}_i$ from the wavevector {\bf q} for a given spin spiral configuration 
fixed at the boundary, we calculate the deviation $\delta q$ of the local spiral state
defined as the angle between the spiral wavevectors ${\bf q}_i$ and {\bf q}, 
e.g. $\cos{(\delta q)} = {\bf q}_i \cdot {\bf q}/(|q||q_i|)$.
Our results for the so-defined spiral deformation for various couplings and boundary
spiral states are summarized in Fig.~\ref{Fig:SpiralWavevector2_qqq_q00}.

We find that the local rearrangement of spins in the vicinity of the
impurity gives rise to a significant deviation of the (angle of the)
local spiral wavevector of $O(1)$, while spins being separated from the
impurity by about one unit cell spacing rearrange themselves in a spiral
state which differs only marginally from the one fixed at the
boundary. This short-range behavior of the spiral deviations is found to
be quite insensitive to the size of the system and the distance from the
fixed boundary configuration; for a more detailed discussion of
finite-size effects see Appendix \ref{appendix:FiniteSize}.
%We find that this result is insensitive to varying the system size as shown in
%the various panels of Fig.~\ref{Fig:SpiralWavevector_qqq}.
%In particular, we
%do not find a striking variation of our results when embedding the impurity into 
%a system of even extent $L$, which places the impurity site rather asymmetrically
%with respect to the fixed boundary spiral.

We further analyze how the pattern of local wavevector deviations
changes as we vary the couplings in the range $1/8 < J_2 / |J_1| < 1/4$.
This is shown in the two panels of Fig.~\ref{Fig:SpiralWavevector2_qqq_q00} 
for fixed boundary spirals pointing in the $1\bar{1}1$ and $100$ directions, respectively.  
We see that the region of significant deformation of the
spiral is in all cases restricted to the very close vicinity of the
impurity, and varies only slightly with varying $J_2/J_1$.  

%-----------------------------
%-----------------------------

\section{Dilute impurities}
\label{sec:dilute-impurities}

\subsection{Ground state with many impurities}

\subsubsection{swiss cheese model}
\label{sec:swiss-cheese-model}

We turn now to the case of many impurities.  We have seen that a single
impurity already breaks the ground state degeneracy of the pure system,
as well as the cubic symmetry of the crystal, thus favoring a
unique state.  However, the (four) different impurity positions 
within the unit cell break the cubic symmetry of the crystal in a different way, 
and hence each favors a different ordered state.  For example, an impurity at one B-site will
favors a spiral with wavevector along the $(111)$ direction, while another
favors a wavevector along the $(11\overline{1})$ direction. In the physical system, 
equal densities of each type of impurity should be simultaneously considered.

%\begin{figure*}
% \centering
%   \includegraphics[width=0.35\linewidth]{./Figures/Imp1.pdf}
%   \hspace{30mm}
%   \includegraphics[width=0.35\linewidth]{./Figures/Imp3.pdf}
% \caption{An impurity which favors a wavevector along the $\overline{1}11$ direction and one which favors a wavevector along the $1\overline{1}1$ direction.
%Magnetic A sites (blue online): original magnetic sites of the lattice.
%Non-magnetic B sites (red online) form a pyrochlore lattice.  The black
%sphere denotes a magnetic impurity on one of the pyrochlore sites, as
%described in Sec.~\ref{XXX}. Black vectors point in the direction of the
%spiral wave vector $q$ of the single impurity.  The six 
%spins closest to the impurity are shaded light blue.   Spins on shaded planes  point in the same direction.
%\textcolor{magenta}{DISCUSSION: what are these supposed to show. What is light vs dark blue? coordinates at arrow heads? Do we need shading of planes? General unhappiness (EG) about figure.}
%\label{fig:Imp1AndImp3}}
%\end{figure*}

Given that the four impurity types favor incompatible orders, what is
the nature of the ground state that emerges here?  We will address this
question in the dilute limit, by which we mean that the impurity density
$n_{\rm{imp}}$ is assumed to be much smaller than $\xi^{-3}$.  One naive
candidate ground state in this limit consists of domains such that
around each defect the spins are close to a spiral with wavevector
favored by that impurity type.  However, this possibility can be
dismissed since such a configuration would necessitate large scale
deviations in wavevector between domains, which we have seen in Sec.\ II
cost a prohibitively large energy.  A more plausible outcome is that the
ground state consists of a \emph{uniform} spiral deformed locally around
the defects, whose wavevector reflects a compromise between the
different impurity types.  Putting it more colloquially, the system
looks like a ``swiss cheese'' (Emmentaler) with the bulk consisting of
an ordered spiral and a set of holes in which the spins are strongly
deformed about each impurity.  In this case, since the energy is the sum
of the energy shifts due to an equal number of each type of impurities,
the ground state wavevector for the many-impurity case minimizes
\begin{equation}
 E({\bf q}) = \frac{1}{4}\sum_{a = 1}^{4} E_a({\bf q}).
 \label{ManyImpurityEnergy}
\end{equation}
Eq.~(\ref{ManyImpurityEnergy}) constitutes a large simplification,
justified by the impurity diluteness---the many-impurity ground state is
determined from an average over single-impurity quantities.  In this
sense, the impurities in this limit act independently.  

The above discussion asserts that the ground state away from the
impurities is essentially undeformed on scales comparable to the
impurity separation and somewhat larger.  This is indeed a consequence
of the assumption of dilute impurities and the locality arguments of
Sec.~\ref{sec:energy-weakly-deform}.  However, this does not rule out
the possibility that small deformations of the spiral on the scale of
the impurity separation could add up on much longer distances to a
larger deviation from long-range spiral order.  We consider this
carefully below.  We find that the {\sl wavevector} of the spiral indeed
remains macroscopically uniform for dilute impurities, even on the
longest scales, with small fluctuations.  This is sufficient to
guarantee the correctness of the energy estimate in
Eq.~(\ref{ManyImpurityEnergy}), and hence correctly predict the
wavevector favored by dilute impurities.  The {\sl phase} of the spiral,
however, fluctuates considerably more, and our arguments suggest that
there may be considerable reduction of the long-range ordered moment of
the spiral by this mechanism.  

To see this, we will construct a ``coarse grained'' energy function for
the system containing many impurities, and consider the stability
against perturbations to a macroscopically uniform spiral.  We use the
parametrization of an arbitrary slowly-varying deviation from a spiral
state with wavevector ${\bf q}_0$ from Sec.~\ref{sec:gener-cons}, in
terms of the fields $\phi$ and $\psi$.  The energy cost in the clean
system for such a deviation is described by Eq.~(\ref{eq:12}).  We must
add to this the impurity energy density,
\begin{equation}
 \label{eq:14}
 {\cal E}_{\rm imp}({\bf q}({\bf r}),{\bf r}) = \sum_a E_a({\bf q}({\bf r})) n_a({\bf r}),
\end{equation}
where the impurity density is
\begin{equation}
 \label{eq:15}
 n_a({\bf r}) = \sum_{{\bf R}_a} \delta({\bf r}-{\bf R}_a),
\end{equation}
and ${\bf R}_a$ are the impurity positions. The impurity density is a
random function.  For long-wavelength properties, the central limit
theorem implies that it is well-characterized by its first few moments.
Taking the impurities to be uniformly and independently distributed over
the system volume with a total average density $x$ (or $x/4$ per
impurity type), we find the mean and two-point correlation
\begin{eqnarray}
 \label{eq:16}
 \overline{n_a({\bf r})} & = & x/4, \\
 \overline{n_a({\bf r})n_b({\bf r}')} -\overline{n_a({\bf
     r})}\;\overline{n_b({\bf r}')}  & = & \frac{x}{4} \delta({\bf 
   r}-{\bf r}') \delta_{ab},
\end{eqnarray}
in the infinite volume limit.  From this, we can evaluate the mean and
second cumulant of the impurity energy density.  The mean is
\begin{equation}
 \label{eq:19}
 \overline{ {\cal E}_{\rm imp}({\bf q},{\bf r}) } = x E({\bf q}).
\end{equation}
This is precisely the energy in Eq.~(\ref{ManyImpurityEnergy}), and is, as expected, linearly
proportional to the impurity concentration $x$.  

As a consequence, the impurity-averaged energy is minimized by the spiral wavevectors that 
minimize $E({\bf q})$. To ascertain the stability of these minima in the impurity distribution
we now turn to analyze fluctuations about the minima of $E({\bf q})$,
parametrized as ${\bf q}={\bf q}_0 + {\boldsymbol \nabla}\phi$ (this is the same slowly varying $\phi$ field from Section~\ref{sec:energy-weakly-deform}).
Consider fluctuations in the impurity energy
\begin{equation}
 \label{eq:23}
 \delta {\cal E}_{\rm imp}({\bf q}({\bf r}),{\bf r}) = {\cal E}_{\rm
   imp}({\bf q}({\bf r}),{\bf r})  - \overline{ {\cal E}_{\rm imp}({\bf
     q}({\bf r}),{\bf r}) },
\end{equation}
and expand to {\sl linear} order in $\phi$:
\begin{equation}
 \label{eq:24}
 \delta {\cal E}_{\rm imp}({\bf q}({\bf r}),{\bf r}) \approx \left[{\cal E}_{\rm
   imp}({\bf q}_0,{\bf r}) - E({\bf q}_0)\right] - {\bf f}_{\rm
   imp}({\bf r}) \cdot {\boldsymbol\nabla}\phi.
\end{equation}

The first term in the brackets is $\phi$-independent and can be
neglected.  The second term represents a ``random force'', given by
\begin{equation}
 \label{eq:25}
 {\bf f}_{\rm imp}({\bf r}) = - \sum_a n_a({\bf r}) {\boldsymbol
 \nabla}_q E_a({\bf q}_0) .
\end{equation}
Since $E({\bf q})$ has a minimum at ${\bf q}_0$, it has vanishing first order derivatives at this point.
This also implies that $\overline{{\bf f}_{\rm imp}({\bf r})} \sim {\boldsymbol \nabla}_q E({\bf q}_0) = 0$.
The second cumulant of the force is
however non-zero:
\begin{equation}
 \label{eq:26}
 \overline{f_{\rm imp}^\mu({\bf r}) f_{\rm imp}^\nu({\bf r}') } = x
 \Delta_{\mu\nu}({\bf q}_0) \delta({\bf r}-{\bf r}'),
\end{equation}
with
\begin{equation}
 \label{eq:27}
 \Delta_{\mu\nu}({\bf q}_0) = \frac{1}{4} \sum_a \frac{\partial E_a({\bf
       q}_0)}{\partial q_\mu} \frac{\partial E_a({\bf
       q}_0)}{\partial q_\nu}.
\end{equation}
$\Delta_{\mu\nu}$ is generally non-zero and positive unless ${\bf q}_0$
is a saddle point for {\sl all} impurity types.  This is not the case
for our problem, but even if it were, it would only further strengthen the
tendency of the system to order.

We are now in a position to consider the full energy function.  Since
$\psi$ does not couple to the impurities, we can neglect it.  The energy
density involving $\phi$ then combines the first terms in
Eq.~(\ref{eq:12}), the mean impurity contribution near a minimum of $E({\bf q})$
to quadratic order in $\phi$
\begin{equation}
 \label{eq:22}
 \overline{ {\cal E}_{\rm imp}({\bf q}({\bf r}),{\bf r}) } \approx x
 E({\bf q}_0) + \frac{x}{2}\frac{\partial^2  E({\bf q}_0)}{\partial
   q_\mu \partial q_\nu} \partial_\mu
 \phi \partial_\nu \phi,
\end{equation}
and the random force from Eq.~(\ref{eq:24}).  Up to an unimportant
additive constant, we find
\begin{eqnarray}
 \label{eq:28}
 {\cal E} & = & \frac{c}{2} (\nabla_\perp \phi)^2 + c' \nabla_\perp\phi
 \nabla_\parallel^2 \phi+
 \frac{c''}{2}(\nabla^2_\parallel \phi)^2 \nonumber \\
&& + \frac{x}{2}\frac{\partial^2  E({\bf q}_0)}{\partial
   q_\mu \partial q_\nu} \partial_\mu
 \phi \partial_\nu \phi -  {\bf f}_{\rm
   imp}({\bf r}) \cdot {\boldsymbol\nabla}\phi.
\end{eqnarray}
To proceed, we note that for dilute impurities (small $x$), the fourth
term in Eq.~(\ref{eq:28}) is much smaller than the first two {\sl
 except} when considering the energy cost for gradients
$\nabla_\parallel \phi$ parallel to the spiral surface, and therefore 
keep only these components.  For simplicity, we will approximate these
components as isotropic, and replace
\begin{equation}
 \label{eq:17}
 \frac{\partial^2  E({\bf q}_0)}{\partial
   q_\mu \partial q_\nu}\partial_\mu
 \phi \partial_\nu \phi  \rightarrow c_{\rm imp} (\nabla_\parallel \phi)^2.
\end{equation}
It is now straightforward to minimize the energy in Eq.~(\ref{eq:28}) in
Fourier space:
\begin{equation}
 \label{eq:18}
 \phi({\bf k}) = \frac{-i {\bf k}\cdot \tilde{f}_{\rm imp}({\bf k})}{c
   k_\perp^2 + c'k_\perp k_\parallel^2+ c'' k_\parallel^4 + x c_{\rm imp} k_\parallel^2}.
\end{equation}
Finally, we can evaluate the local variance of the wavevector $\delta
{\bf q}= {\boldsymbol\nabla\phi}$:
\begin{eqnarray}
 \label{eq:20}
 \overline{\delta{\bf q}({\bf r})^2} & = & \int_{\bf k} k^2
 \overline{\phi({\bf k})
   \phi(-{\bf k})} \\ & = &  x \Delta_{\mu\nu} \int_{\bf k} \frac{k^2 k_\mu k_\nu}{(c
   k_\perp^2 + c'' k_\parallel^4 + x c_{\rm imp} k_\parallel^2)^2 -
   (c')^2 k_\perp^2 k_\parallel^4}.\nonumber
\end{eqnarray}
To estimate the integral for small $x$, we note the denominator of the
integrand vanishes more rapidly with $k_\parallel$ than with $k_\perp$,
and hence the largest terms will be those in which the momenta in the
numerator are taken in the $k_\parallel$ directions.  Hence, up to
angular factors which do not affect the scaling with $x$, we estimate 
\begin{equation}
 \label{eq:21}
  \overline{|\delta{\bf q}({\bf r})|^2} \sim x|\Delta|
 \int d^2k_\parallel dk_\perp \frac{k_\parallel^4}{(c
   k_\perp^2 + c'' k_\parallel^4 + x c_{\rm imp} k_\parallel^2)^2-
   (c')^2 k_\perp^2 k_\parallel^4}.
\end{equation}
The integral over $k_\perp$ can be performed directly to obtain
\begin{equation}
 \label{eq:30}
  \overline{|\delta{\bf q}({\bf r})|^2} \sim x|\Delta|
\frac{1}{\sqrt{c}} \int_0^\Lambda dk_\parallel \frac{k_\parallel^2}{(
  \tilde{c}^{''} k_\parallel^2 + x c_{\rm imp} )^{1/2}(
  c^{''} k_\parallel^2 + x c_{\rm imp} ) },
\end{equation}
where $\tilde{c}'' = c'' - (c')^2/(4c)$, and we have introduced the
radial momentum coordinate $k_\parallel$ and introduced a high momentum
(short distance) cut-off $\Lambda$. The integral is readily seen to be
logarithmically divergent for small $x$, hence
\begin{equation}
 \label{eq:29}
 \overline{|\delta{\bf q}({\bf r})|^2} \sim \frac{|\Delta| x}{\sqrt{c}} \ln(1/x).
\end{equation}
In the limit $x \rightarrow 0$ the fluctuations of the wavevector vanish,
and therefore fluctuations never diverge. 
The wavevector is indeed expected to remain uniform over the entire system, 
with only small fluctuations for small $x$.

A more subtle question concerns the deformation of the {\sl phase}
$\phi$ rather than the wavevector, because two well-separated regions of
the sample can become arbitrarily out of phase as small deformations of
the spiral accumulate between them. A similar analysis to above gives
\begin{eqnarray}
 \label{eq:33}
 \overline{|\phi({\bf r})|^2} & = & \int_{\bf k}
 \overline{\phi({\bf k})
   \phi(-{\bf k})} \\ & \sim &  x|\Delta|
\frac{1}{\sqrt{c}} \int_0^\Lambda dk_\parallel \frac{1}{(
  \tilde{c}^{''} k_\parallel^2 + x c_{\rm imp} )^{1/2}(
  c^{''} k_\parallel^2 + x c_{\rm imp} ) }. \nonumber
\end{eqnarray}
The integral in this case is much more singular.  For small $x$ it is
dominated by small $k_\parallel$ and independent of $\Lambda$.  By
rescaling, one finds it is proportional to $1/x$, {\sl canceling} the
$x$ dependence of the prefactor:
\begin{equation}
 \label{eq:31}
 \overline{|\phi({\bf r})|^2} \gtrsim \frac{|\Delta|}{c_{\rm imp} \sqrt{c c''}}.
\end{equation}

Because Eq.~(\ref{eq:31}) is independent of $x$, there is no particular
reduction of the spatial variations of the spiral phase for dilute
impurities.  This is symptomatic of the ``softness'' of the degenerate
spiral manifold.  

Inspecting both Eqs.~(\ref{eq:31},\ref{eq:29}), we see that, although
fluctuations do not become large for small $x$, they {\sl do} become
large for small $c$.  Since $c$ vanishes on approaching the Lifshitz
point $J_2/J_1=1/8$, we expect that the spiral ordering should become
unstable to impurity deformations in the neighborhood of this part of
the phase diagram.  We return to this point in the Discussion.

%% Numerical results %%

\subsubsection{Numerical results}

We have argued above that dilute impurities basically act independently 
of each other and that they favor a unique ground-state wavevector which 
minimizes the energy  $E({\bf q})$. 
As a consequence, it is straightforward to estimate the impurity average $E({\bf q})$
from our numerical calculations of $E_a({\bf q})$ for a single impurity 
in section \ref{sec:numerical-results-single-impurity}. 
Our results for the impurity averaged energies $E({\bf q})$ are summarized in the 
bottom row of Fig.~\ref{Fig:ImpuritySelection}.
Note that while $E_a({\bf q})$ does not have the full point group symmetry of 
the lattice, cubic symmetry is restored when calculating the average 
$E({\bf q})$.

In particular, our numerical results allow us to determine the direction of the 
long-distance spiral wavevector favored by an ensemble of dilute impurities.
For couplings $1/8 < J_2/|J_1| < 1/4$, multiple defects favor a long-distance spiral
wavevector residing on the spiral surface along one of the $100$ directions.  
For couplings $J_2/|J_1| > 1/4$ where the spiral surface develops `holes' centered
around the $111$ directions, we find that also the long-distance spiral wavevector 
favored by an ensemble of dilute impurities first jumps to the $1\bar{1}1^*$ direction
for $1/4<J_2/|J_1| \lesssim 0.30$, and then continuously moves to the $100^*$ directions
for $J_2/|J_1| \gtrsim 0.30$ as illustrated in Fig.~\ref{Fig:ImpuritySelection}.

\subsection{Interplay between impurity and entropic effects}

We have argued that at zero temperature, dilute impurities lift the spiral degeneracy inherent in
the pure system, generating ``order by quenched disorder''.  As discussed above and in
Ref.~\onlinecite{nature07}, entropy provides another degeneracy lifting mechanism at finite
temperature via ``thermal order by disorder''.  The interplay between these mechanisms leads to
interesting physics as we will now discuss.  In particular, over a wide range of $J_2/J_1$,
disorder and thermal fluctuations favor decidedly different ordered states; \emph{e.g.}, for
$J_2/J_1 = 0.2$, thermal fluctuations favor the 111 directions while impurities prefer the 100
directions.  In such cases, since entropic corrections giving rise to thermal order by disorder
vanish as $T\rightarrow 0$, the system is expected to exhibit multi-stage ordering, from an
impurity-driven phase at the lowest $T$ to an entropically stabilized phase at moderate $T$ to a
disordered paramagnet at still higher $T$.  As an aside, we note that other interactions beyond
those considered in our model and/or quantum fluctuations can compete with impurity effects at
low $T$, but may similarly lead to multiple phase transitions.  If the energetic corrections
coming from impurity or other effects are too large, however, then the entropically stabilized
phase will be removed, leaving a single ordered state.

Another interesting effect arising from the interplay between entropy and disorder, which can be
probed experimentally, pertains to the shift in transition temperature $T_c$ at which the system
first orders.  Roughly, should entropy and disorder favor the same state, then $T_c$ is expected
to be enhanced relative to the pure system; otherwise a reduction is anticipated.  To estimate
this shift, we note that the transition is first-order and that at $T_c$ the free energies for
the paramagnet and the ordered phase must equal,
\begin{equation}
 f_{sp}\left(T_{c}\right)+x\delta F_{sp}\left(T_{c}\right)=f_{PM}\left(T_{c}\right)+x\delta F_{PM}\left(T_{c}\right).
\end{equation}
Here, $x$ is the impurity concentration, $f_{sp}$ and $f_{PM}$ are the free energies for a clean
system in the spiral phase and paramagnet, respectively, and $x\delta F_{sp}$ and $x\delta
F_{PM}$ are the corresponding changes in free energy due to the impurities. {\it For a well
defined thermal order-by-disorder phase}, an approximate derivation (see Appendix~\ref{append:computation-Tc-shift})
yields the following result
\begin{equation}
T_c - T_c^*
= 
T_c^* x 
\frac{ \left[E(q)\right]_S - E({\bf q}_0) }{l^*}
\; ,\label{eq:43}
\end{equation}
where $T_{c}^{*}$ is the (upper) ordering temperature for the clean system, $l^*$ is the latent
heat density to go from the ordered phase to the paramagnetic one in the clean crystal, $S$ is
the degeneracy (spiral) surface and ${\bf q}_0$ is the momentum favored
by thermal order-by-disorder in the clean system.  We defined the surface average
\begin{equation}
  \label{eq:41}
   \left[E(q)\right]_S = \frac{\int_{{\bf q} \in S} d{\bf q} E({\bf q})}{\int_{{\bf q} \in S} d{\bf q}} .
\end{equation}

\begin{figure}[t]
\includegraphics[width=0.8\columnwidth]{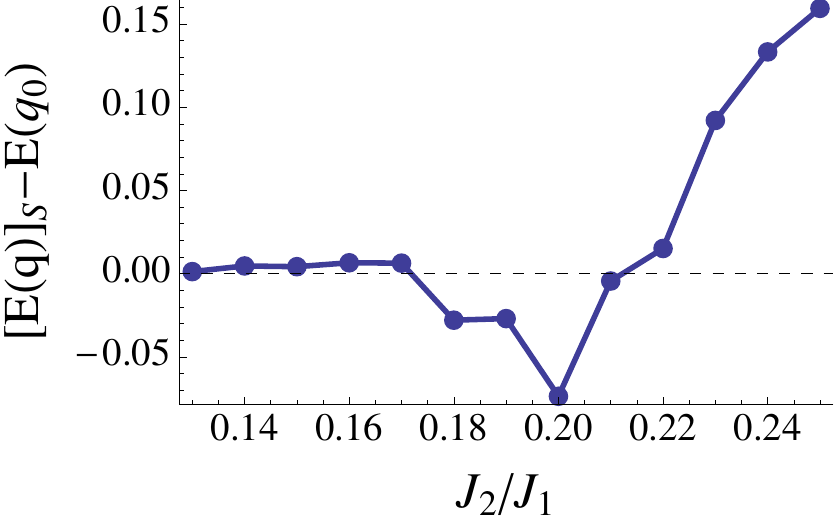}
\caption{Plot of $\left[E(q)\right]_S - E({\bf q}_0)$ versus
  frustration $J_2/J_1$.  The shift of $T_c$ for the order-by-disorder
phase per impurity is proportional to this quantity (see Eq.~\eqref{eq:43}).  }
\label{fig:Tc-shift}
\end{figure}

The quantities determining the $T_c$ shift in
Eq.~\eqref{eq:43} can be extracted from numerics.  We find that the
latent heat $l^*$ is roughly independent of $J_2/J_1$.  However, there
is significant dependence of the numerator in Eq.~\eqref{eq:43} on
this ratio.  This is plotted in Fig.~\ref{fig:Tc-shift}. The
$T_c$ tracks this quantity, and is thus sensitive to the degree of frustration.  

%-----------------------------
%-----------------------------

\section{Discussion}
\label{sec:discussion}

In this manuscript we have explored the effect of dilute impurities on
the $J_1 - J_2$ model on the diamond lattice.   General considerations led
us to conjecture that impurities may provide a mechanism for ground
state degeneracy breaking.   We established that, under rather general
conditions, even highly frustrated magnets are induced to {\sl order} by
low concentrations of impurities.  Moreover, the mechanism and
energetics of this ordering was explained in terms of a simple ``swiss
cheese'' picture.  To expose the mechanism in more detail, we
considered a very specific impurity model, namely B site magnetic ions being
added to the system, and confirmed the general structure of the
impurity-induced ordering by numerical and analytical means in this
situation.  

Let us briefly discuss this picture in relation to \cao\ and \mss, the
two A-site magnetic spinels exhibiting the largest frustration
parameters without the complications of orbital degeneracy.  Disorder in
the form of inversion -- A and B site atoms interchanging with one
another -- is prevalent in many spinels, including these, at the level
of at least a few percent.  One intriguing feature of the measurements
on these materials is the observation of glassy freezing in \cao,
but not in \mss, despite comparable levels of inversion in the two
materials.  This suggests that \cao\ is more sensitive to defects than
\mss, and our results corroborate this hypothesis.  Theoretically, we
argued that, {\sl generically}, the effect of {\sl sufficiently dilute}
impurities is to induce order, not a spin glass.  For the ordered state
to be stable, we argued that: (1) the impurity ``halos'' should not
overlap, and (2) the fluctuations in wavevector induced by the
randomness of the impurity positions should be small.  In
Sec.~\ref{sec:swiss-cheese-model}, we saw that the second criteria is
highly sensitive to the magnitude of the stiffness $c$, the fluctuations
becoming large as $c$ decreases.  In the diamond lattice
antiferromagnets, the stiffness $c$ actually {\sl vanishes} on
approaching the Lifshitz point $J_2/J_1=1/8$.  Prior investigations
concluded that in fact \cao\ has exchange parameters close to this
point, while in \mss, $J_2/J_1 \approx 0.85$,\cite{nature07} where $c$ is not small.
Thus we suggest that the freezing behavior in \cao\ may be understood
as arising from proximity to the Lifshitz point.  It may be interesting
to directly study disorder physics in this region by field theoretic
methods in the future.

We would like to emphasize the generality of this argument.  The only
assumption is that the impurity positions are not strongly correlated,
but otherwise this conclusion is independent of the type of defects.
Indeed, we do not maintain any direct relevance of the specific impurity
modeled studied in the numerical portions of this paper to the A-site
spinels.  For \cao, the existence of magnetic ions on the B sites is
probably suspect, as inverted Co$^{+3}$ on the B sites would be expected
to have a non-magnetic ground state.  However, the expected spin
``vacancies'' induced by Al atoms on the A sites would lead to the same
general conclusions.  What {\sl would} require a more appropriate
microscopic  model would be an estimate of the size of the
region of deformed spins around an impurity.  

Very recent experiments have greatly clarified the situation in \cao.
Through a careful study of elastic and inelastic neutron scattering
high quality single crystal, MacDougall {\sl et al.} \cite{2011arXiv1103.0049M} 
have argued that the freezing
transition in \cao\ signals an ``arrested'' first order transition in
which the sample breaks up into antiferromagnetic domains.  These domains
are evidenced by a substantial Lorentzian-squared component to the
elastic scattering.  Moreover, below the freezing temperature
spin-wave excitations were observed, a fit of which determined
$J_2/J_1 \approx 0.1$.  This parameter ratio takes \cao\ close to the Lifshitz point but $0.1<1/8$, so the
commensurate N\'eel state would be expected at low temperature.  The
first order nature of the transition is consistent with theoretical
expectations based on the order-by-disorder mechanism.\cite{nature07}
Given these exchange parameters, the detailed analysis of this paper
does not directly apply, since we have assumed $J_2/J_1 > 1/8$ and
focused on spiral ground states.  However, arguments very similar to those we applied
here to show a strong sensitivity to impurities close to the
Lifshitz point on the spiral side also imply a similar sensitivity
close to the Lifshitz point on the N\'eel side.  Thus the findings are
quite consistent with the general reasoning espoused here.  

We conclude by describing an interesting feature of our numerical
simulations, which might be of interest in future theoretical and
experimental studies.  We found that, while the ground states of the
pure system are coplanar, the spin configuration around the impurity
might acquire a sizable out-of-plane spin component, i.e. a spin
component orthogonal to the plane in which the spin
spiral state lies at long distances away from the impurity.  This is in
particular true for those spirals with wavevectors $\mathbf{q}$ such that their energies $E_1({\bf q})$ (see Eq.\eqref{eq:en-corresp}) are far away from the overall minimum.  It is possible
that, collectively, impurities might therefore induce non-coplanar spin
ordering.  Such non-coplanar order is relatively rare, and interesting
insofar as it can induce non-trivial Berry phases, related to anomalous
Hall effects in conducting systems.

\section*{Acknowledgments}
We would like to thank Leo Radzihovsky for extensive discussions during
the prehistory of this project, and apologize for his wasted time.  Our
numerical simulations were based on the classical Monte Carlo code of
the ALPS libraries \cite{ALPS}.  L.B. was supported by the Packard
Foundation and National Science Foundation through grants DMR-0804564
and PHY05-51164.  J.A. acknowledges support from the National Science Foundation through grant DMR-1055522.

\appendix

\section{Definition of local wavevector}
\label{sec:defin-local-wavev}

Here we describe in detail how the local spiral wavevector is defined on
the lattice, as used in
Sec.~\ref{sec:numerical-results-single-impurity}\ and
Figs.~\ref{Fig:SpiralWavevector2_qqq_q00} and \ref{Fig:SpiralWavevector_qqq}.
Depending on the unperturbed spiral wavevector ${\bf q}$ taken at
infinity we consider distinct sets of three neighboring sites out of the
12 second-neighbor sites which are nearest neighbors on the identical
(fcc) sublattice.  In particular, for ${\bf q}$ pointing in the
$1\overline{1}1$ direction we consider two sets of vectors $\{ {\bf
  r}_{ij} \}$, namely $P = \{(1/2, -1/2, 0); (1/2, 0, 1/2); (0, -1/2,
1/2)\}$ and $Q = \{(-1/2, -1/2, 0); ( 1/2, 0, -1/2); (0, -1/2, -1/2)
\}$.  For ${\bf q}$ pointing in the $100$ direction we consider two
alternative sets of vectors $\{ {\bf r}_{ij} \}$, namely $P' = \{(1/2,
1/2, 0); (1/2, -1/2, 0); (1/2, 0, 1/2)\}$ and $Q' = \{(1/2, 1/2, 0); (
1/2, 0, 1/2); (1/2, 0, -1/2) \}$.  We place the local wavevector ${\bf
  q}_i$ at position ${\bf r}_i - \frac{1}{4} \sum_{j=1}^{3} {\bf
  r}_{ij}$, which is always located inside the (convex) manifold spanned
by the four spins.

\section{Symmetries}

In this appendix, we give explicit expressions for the symmetry
transformations and their effects, within our conventions for the spinel
lattice.  The space group is generated by the
following operations:
\begin{enumerate}
\item A three-fold rotation about the $(1,1,1)$ axis: 
  \begin{equation}
    \label{eq:3fold}
    T_1: (x,y,z)\longrightarrow(z,x,y).  \end{equation}
\item A two-fold rotation about the $(0,0,1)$ axis: 
  \begin{equation}
    \label{eq:2fold}
    T_2: (x,y,z)\longrightarrow (-x,-y,z).
  \end{equation}
\item Reflection through a $(1,-1,0)$ plane: 
  \begin{equation}
    \label{eq:reflec}
    T_3: (x,y,z)\longrightarrow(y,x,z)    .
  \end{equation}
\item Inversion: 
  \begin{equation}
    \label{eq:invers}
    T_4: (x,y,z)\longrightarrow(\tfrac{1}{4}-x,\tfrac{1}{4}-y,\tfrac{1}{4}-z).
  \end{equation}
\end{enumerate}
%The symmetries are such that, numerically, the study of the boundary conditions can be limited to
%a fraction of the spiral surface, for example, for impurity of type 3 to
%$\mathbf{k}=(k_x,k_y,k_z)$'s with:
%\begin{equation}
%-k_{x}+k_{y}+k_{z}\geq0\qquad k_x+k_y\geq0\qquad k_x+k_z\leq0
%\end{equation}

We define the following four impurity positions, ${\bf u}_a$ ($a=1,2,3,4$) modulo
Bravais lattice transformations:
\begin{eqnarray}
  \label{eq:imp-pos}
  {\bf u}_1 & = (3/8,5/8,3/8) \qquad {\bf u}_2 & = (3/8,3/8,5/8), \\
  {\bf u}_3 & = (5/8,3/8,3/8) \qquad {\bf u}_4 & = (5/8,5/8,5/8). \nonumber
\end{eqnarray}

These positions are mapped into one another by the four space group
generators.  Corresponding to each of these generators is an associated
linear transformation in reciprocal space.  This transformation of
wavevectors is {\sl identical} to the transformation of real space
coordinates {\sl except} that translational components of the
transformation are dropped.  That is, if the coordinates transform
according to ${\bf r} \rightarrow O {\bf r} + {\bf a}$ ($O$ is an O(3)
matrix), then the corresponding momentum transformation is just ${\bf q}
\rightarrow O {\bf q}$.  

As a consequence, any given impurity position may be mapped to the other
three by such an O(3) operation.  One finds that (up to Bravais lattice
vectors), the impurity positions transform according to
\begin{equation}
  \label{eq:formal-trans}
  T_a {\bf u}_b =  {\bf u}_{c(b,a)},
\end{equation}
where $c(a,b)$ can be represented as the matrix
\begin{equation}
  \label{eq:trans}
  c(a,b) = \left(\begin{array}{cccc}
2 & 3 & 3 & 1 \\
3 & 4 & 2 & 2 \\
1 & 1 & 1 & 3 \\
4 & 2 & 4 & 4\end{array}\right),
\end{equation}
where $a$ and $b$ specify the row and column of the matrix,
respectively.  We see from this that, for instance, an impurity on
position 4 retains the symmetries generated by $T_1$, $T_3$ and $T_4$,
but not $T_2$.

Moreover,  we observe that each impurity position can be mapped to
position 1 in the following way:
\begin{eqnarray}
  \label{eq:imp-mapping}
  {\bf u}_1 & = & T_1 \circ T_1 \, {\bf u}_2, \\
  {\bf u}_1 & = & T_1 \, {\bf u}_3, \\
  {\bf u}_1 & = & T_1 \circ T_1 \circ T_2 \, {\bf u}_4.
\end{eqnarray}
This allows one to calculate the
energies $E_a({\bf q})$ with $a=2,3,4$ from $E_1({\bf q}')$ with an
appropriate ${\bf q}'$.  Specifically
\begin{eqnarray}
  \label{eq:en-corresp}
  E_2(q_x,q_y,q_z) & = & E_1(q_y,q_z,q_x),
  \nonumber \\
  E_3(q_x,q_y,q_z) & = & E_1(q_z,q_x,q_y),
  \nonumber \\
  E_4(q_x,q_y,q_z) & = & E_1(q_y,-q_z,q_x) .
\end{eqnarray}
Therefore, the average energy can be written as
\begin{eqnarray}
  \label{eq:averageEn-E1}
  E(q_x,q_y,q_z) &=& \frac{1}{4}\Big[ E_1(q_x,q_y,q_z) +
  E_1(q_y,q_z,q_x) \nonumber \\
  && + E_1(q_z,q_x,q_y) +E_1(-q_y,q_z,-q_x) \Big].\nonumber \\
\end{eqnarray}
We note that, taking into account the subgroup of the full space group
which leaves position 1 invariant, the first impurity energy obeys
\begin{eqnarray}
  \label{eq:equivpointsE1}
 && E_1(-q_y,-q_z,q_x) =E_1(q_z,q_y,q_x)=E_1(q_y,q_z,-q_x)  \\
&  & =  E_1(-q_z,-q_y,-q_x)=E_1(q_x,-q_z,-q_y)=E_1(q_z,-q_x,-q_y)\nonumber \\
 &&=  E_1(-q_x,q_z,q_y)=E_1(-q_z,q_x,q_y)=E_1(q_x,q_y,q_z)\nonumber \\
&& = E_1(-q_y,-q_x,q_z)=E_1(-q_x,-q_y,-q_z)=E_1(q_y,q_x,-q_z).\nonumber
\end{eqnarray}
The average energy, by construction, has the {\sl full} cubic space
group symmetry, i.e.
\begin{equation}
  \label{eq:equivpoints}
  E(q_x,q_y,q_z) = E(s_a q_a, s_b q_b, s_c q_c),
\end{equation}
where $s_a,s_b,s_c = \pm 1$ and $(q_a,q_b,q_c)$ is an arbitrary 
permutation of $q_x,q_y,q_z$.  As a consequence, $1/48$th of the solid
angle in ${\bf q}$ space is enough to recover the full function $E({\bf
  q})$.  We therefore carry out numerical simulations only for such a
section, which we choose, arbitrarily, to be the one defined by:
\begin{eqnarray}
  \label{eq:twelfth}
&&  (q_x>0) \wedge (q_y>0) \wedge (q_z>0) \wedge (q_x>q_y) \wedge (q_x <
  q_z)\nonumber \,. \\
\end{eqnarray}
All points defined by Eq.~\eqref{eq:twelfth} are inequivalent to one another,
and conversely, can used to generate $E({\bf q})$ for an arbitrary point
using Eq.~\eqref{eq:equivpoints}.

\section{Locality of spin deformation and finite-size effects}
\label{appendix:FiniteSize}

\begin{figure}[t]
\includegraphics[width=\columnwidth]{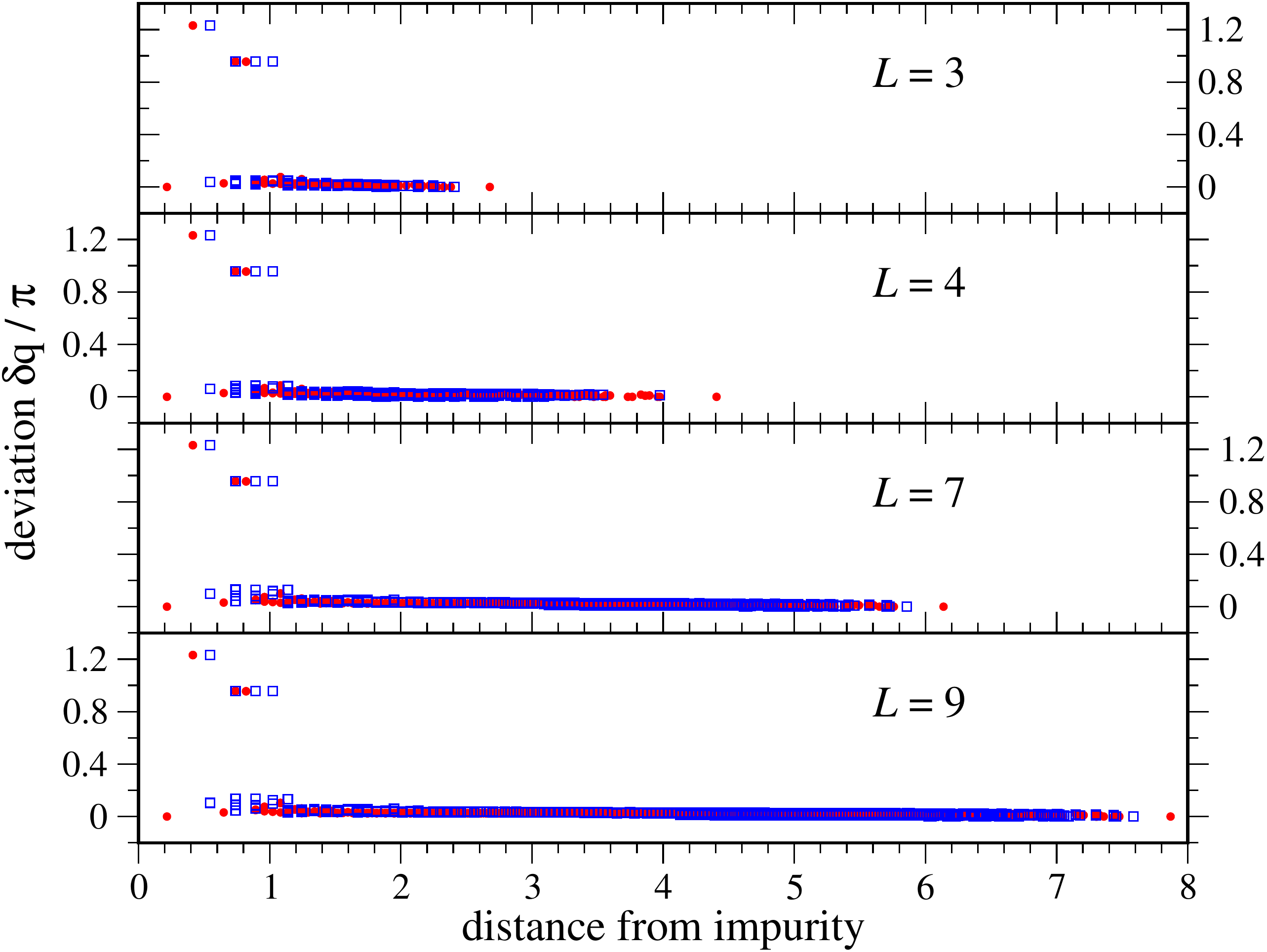}
\caption{
 Finite size effects: Deviation of the local spiral wavevector ${\bf q}_i$ from the one 
 of the spiral state at the boundary $\bf q$ for systems of varying
 size $N = 8 \times L^3$.
 For the chosen couplings $J_2/|J_1|=0.2$ the spiral wavevector ${\bf q}$ 
 points along  the $1\bar{1}1$ direction with length $|{\bf q}|=0.71\pi$.
 The symbols correspond to two different sets of neighboring spins, 
 $P$ (circles) and $Q$ (boxes), used to calculate the local spiral 
 wavevectors, details are given in the text.
}
\label{Fig:SpiralWavevector_qqq}
\end{figure}

Our numerical simulations of the B-site impurity model indicate that the deformation 
of the spin spiral state in the vicinity of the impurity is limited to a small numbers of spins
in the unit cell around the impurity. 
Our numerical calculations are performed in an $L\times L\times L$ simulation cube 
embedded in a larger cube of extent $L+1$, where the spins in the `boundary cube' are fixed
to a particular spin spiral state. One might thus wonder whether the locality of the spin spiral
deformation originates from the impurity physics, as opposed to artifacts due to the fixed 
boundary conditions. To exclude the latter we have calculated the spiral deviation for different
system sizes $L$ and positioning of the impurity site, as summarized in Fig.~\ref{Fig:SpiralWavevector_qqq} where we fix the boundary spiral wavevector to the $1\bar{1}1$ direction, which  minimizes $E_1({\bf q})$ for the chosen ratio of couplings $J_2/|J_1|=0.2$.
We find that the deviation is insensitive to varying the system size as shown in
the various panels.
Further, we also do not find a striking change of our results when embedding the impurity into 
a system of even extent $L=4$ (second panel from top in Fig.~\ref{Fig:SpiralWavevector_qqq}), which places the impurity site rather asymmetrically with respect to the fixed boundary spiral.

%-----------------------------
%-----------------------------

\section{Derivation of the transition temperature shift}
\label{append:computation-Tc-shift}

In this appendix we consider the transition between the high temperature paramagnetic phase, 
and an ordered phase where the spins order in a spiral configuration 
with a wavevector selected by entropy. We explore how adding impurities shifts the transition temperature,
assuming that the impurities themselves do not change the nature of the ordered state.
Quantities in the clean limit are denoted by a star.

For a first-order phase transition, at the transition temperature:
\begin{equation}
\begin{array}{l}
f_{sp}\left(T_{c}\right) + x\delta F_{sp}\left(T_{c}\right)=f_{PM}\left(T_{c}\right) + x\delta
F_{PM}\left(T_{c}\right)\\
f_{sp}\left(T_{c}^{*}\right) =f_{PM}\left(T_{c}^{*}\right),
\end{array}
\end{equation}
where $f_{sp,PM}$ are the free energy densities of the spiral phase and paramagnetic phase, respectively,
and similarly $\delta F_{sp,PM}$ refer to the free energy density corrections when impurities are included.
A small impurity concentration will only slightly shift the transition temperature, and so we expand
the free energy density to first order in the temperature shift
\begin{equation}
\begin{split}
f(T_c) \approx &
f(T_c^*) + \frac{\partial f(T_c^*)}{\partial T} (T_c - T_c^*)
\\ = &
f(T_c^*) + s(T_c^*) (T_c - T_c^*) 
\\ = & 
f(T_c^*) + \left( f(T_c^*) - \epsilon \right) \left( \frac{T_c}{T_c^*} - 1 \right)
\; ,
\end{split}
\end{equation}
where $s(T)$ is entropy, and $\epsilon$ is the energy density.
From this we find
\begin{equation}\label{diff_eq}
\left( \epsilon_{PM} - \epsilon_{sp} \right) \left( \frac{T_c}{T_c^*} - 1 \right)
=
x \left( \delta F_{PM}(T_{c}) - \delta F_{sp}(T_{c}) \right)
\; .
\end{equation}

Now we turn to estimate the free energy densities $\delta F$,
which can be estimated from $F = -T \log{Tr\left[ e^{-\beta H} \right]}$,
when varying the Hamiltonian by a small term $x \delta H$. This will yield
a small change in the free energy
\begin{equation}
\begin{split} &
\delta F = -T \log{\left[\frac{Tr\left[ e^{-\beta (H + \delta H)} \right]}{Tr\left[ e^{-\beta H} \right]}\right]} 
\\ &
\approx
- T \log{\left[1- \beta \langle \delta H \rangle \right]} 
\approx
+ \langle \delta H \rangle 
\; ,
\end{split}
\end{equation}
where the angle brackets denote a thermal average.
Each impurity will contribute a term of the form of \eqref{eq:13} to $\delta H$.
Next we estimate the energy thermal average in each phase. In the ordered phase,
the system remains mostly in the ground state configuration, and so we estimate
$\delta F_{sp} \approx E({\bf q}_0)$, where $E({\bf q})$ is the same as in \eqref{eq:14},
and ${\bf q}_0$ is the spiral wavevector. In the paramagnetic phase, close to the transition temperature,
the system thermally fluctuates mostly amongst the different spiral states
(this has been shown explicitly for the clean system in Ref.~\onlinecite{nature07})
and so we estimate
$\delta F_{PM} \approx \int_{{\bf q} \in S} d{\bf q} E({\bf q})/(\int_{{\bf q} \in S} d{\bf q})$
where $S$ is the spiral surface. We find therefore
\begin{equation}
\left( \epsilon_{PM} - \epsilon_{sp} \right) \left( \frac{T_c}{T_c^*} - 1 \right)
=
x \left( \frac{\int_{{\bf q} \in S} d{\bf q} E({\bf q})}{(\int_{{\bf q} \in S} d{\bf q})} - E({\bf q}_0) \right)
\; ,
\end{equation}
and finally
\begin{equation}
T_c - T_c^*
= 
\frac{T_c^* x }{l^*}
\left( \frac{\int_{{\bf q} \in S} d{\bf q} E({\bf q})}{(\int_{{\bf q} \in S} d{\bf q})} - E({\bf q}_0) \right)
\; ,
\end{equation}
where $l^* = \left( \epsilon_{PM} - \epsilon_{sp} \right)$
is the latent heat density to go from the order-by-disorder phase to the paramagnetic
phase in a clean system.

\bibliography{Diamond_AFM}

\end{document}